\documentclass[aps,pr,twocolumn,superscriptaddress,showpacs]{revtex4}
\usepackage{amssymb}
\usepackage{epsfig}
\usepackage{amsmath}
\usepackage{times}
\usepackage{graphicx}
\usepackage{amsbsy}
\usepackage{latexsym}
\usepackage{color}
\usepackage{psfrag}
\usepackage[normalem]{ulem}
\usepackage{float}

\begin{document}
	
\title {Fluctuation induced intermittent transitions between distinct rhythms in balanced excitatory-inhibitory spiking networks}

\author{Xiyun Zhang}
\email{xiyunzhang@jnu.edu.cn}
\affiliation{Department of Physics, Jinan University, Guangzhou, Guangdong 510632, China}

\author{Bojun Wang}
\affiliation{Department of Physics, Jinan University, Guangzhou, Guangdong 510632, China}

\author{Hongjie Bi}
\email{hongjiebi@gmail.com}
\affiliation{School of Chemical Biology and Biotechnology, Peking University Shenzhen Graduate School, Shenzhen, Guangdong 518055, China}

\date{\today}

\begin{abstract}
Intermittent transitions, associated with critical dynamics and characterized by power-law distributions, are commonly observed during sleep. These critical behaviors are evident at the microscopic level through neuronal avalanches and at the macroscopic level through transitions between sleep stages. To clarify these empirical observations, models grounded in statistical physics have been proposed. At the mesoscopic level of cortical activity, critical behavior is indicated by the intermittent transitions between various cortical rhythms. For instance, empirical investigations utilizing EEG data from rats have identified intermittent transitions between $\delta$ and $\theta$ rhythms, with the duration of $\theta$ rhythm exhibiting a power-law distribution. However, a dynamic model to account for this phenomenon is currently absent. In this study, we introduce a network of sparsely coupled excitatory and inhibitory populations of quadratic integrate-and-fire (QIF) neurons to demonstrate that intermittent transitions can emerge from the intrinsic fluctuations of a finite-sized system, particularly when the system is positioned near a Hopf bifurcation point, which is a critical point. The resulting power-law distributions and exponents are consistent with empirical observations. Additionally, we illustrate how modifications in network connectivity can affect the power-law exponent by influencing the attractivity and oscillation frequency of the stable limit cycle. Our findings, interpreted through the fundamental dynamics of neuronal networks, provide a plausible mechanism for the generation of intermittent transitions between cortical rhythms, in alignment with the power-law distributions documented in empirical researches.  
	  
\end{abstract}
\maketitle

\section{Introduction}

The intricate functions and various states of the brain are rooted in the collaboration and interaction of neuron assemblies across different spatial and temporal scales. One of the most intriguing collective phenomena arising from the joint activity of large population of neurons is cortical rhythms, which are consistently associated with various brain functions and physiological states \cite{Buzsaki:2004,Buzsaki:2012}. Spontaneous transitions between different physiological states are commonly observed in the brain during resting states \cite{Rechtschaffen:1968,Massimini:2005,Huo:2022}, during task performance \cite{Buzsaki:2006}, and in various pathological conditions \cite{Stam:2007,Chavez:2010}. These transition behaviors typically differ from those observed in equilibrium systems \cite{Brown:2012,Halasz:1998,Lo:2002,Hirshkowitz:2002,Lo:2004,Saper:2005}. Instead, they resembles the intermittent behaviors found in systems that display non-equilibrium dynamics at criticality. Such critical dynamics are often indicated by a power-law distribution of the durations of activated states and are believed to underlie the optimal processing and computational capabilities of neural networks \cite{beggs:2008,beggs:2012,shew:2013}.  

Sleep is crucial for maintaining brain function, making it a vital physiological behavior in vertebrates. It is not merely a passive state; rather, it is an active and complex process that alternates across different sleep stages, each with distinct physiological characteristics \cite{Rechtschaffen:1968}. Numerous studies indicate that these transitions are intermittent and  exhibit critical behavior at various scales, characterized by power-law distributions related to wakefulness and arousal (the active state). At the macroscopic level, which encompasses the entire human body, this critical behavior is evident in the intermittent shifts between sleep stages, with a power-law distribution observed in the duration of wakefulness and arousal \cite{Lo:2002,Lo:2004}. To explore this phenomenon, models based on diffusion processes and state propagation have been proposed \cite{Huo:2022,Dvir:2018}. At the microscopic level, concerning the spiking dynamics of neuronal populations, critical behavior is linked to neuronal avalanches, which are reflected in power-law distributions for the sizes and durations of these avalanches \cite{Bocaccio:2019,Scarpetta:2023}. Consequently, statistical models have also been developed to explain these findings \cite{Lombardi:2023}.

In the mesoscopic scale, which focuses on cortical activities, critical behaviors are also observed. For instance, research utilizing EEG recordings from rodents has shown that brain dynamics frequently switch between states with relatively stronger $\delta$ waves (indicative of sleep, quiet states) and $\theta$ waves (indicative of wakefulness, activated states). The durations of these states follow different statistical distributions, with $\theta$ wave durations exhibiting a power-law distribution and $\delta$ wave durations following a logarithmic distribution \cite{Wang:2019,Huo:2024,Lombardi:2020}. Additionally, the power-law exponent for $\theta$ wave duration changes when sleep disorders are induced \cite{Huo:2024,Lombardi:2020}. However, the dynamic mechanisms driving these observations at this scale are still largely unexamined. To address this, we employ a balanced excitatory-inhibitory network composed of Quadratic Integrate and Fire (QIF) neurons to study the intermittent transitions between different oscillation rhythms. Our findings suggest that this intermittent behavior can arise from intrinsic fluctuations in a finite-size system when parameters are close to the Hopf bifurcation point (critical point). Furthermore, the duration distributions of the two alternating rhythms show similarities to those found in empirical data, and the power-law exponent for the $\theta$ rhythm can be changed by network connectivity.

\section{Networked QIF Model}
In this study, we examine sparsely coupled excitatory and inhibitory populations of QIF neurons, a framework commonly employed in the modeling of cortical rhythms \cite{Ermentrout:1986,diVolo:2018,Bi:2020,Bi:2021}. The system consists of \(N^{(e)}\) excitatory neurons and \(N^{(i)}\) inhibitory neurons. The dynamical equations governing the membrane potentials \(v_j^{(e)}\) and \(v_j^{(i)}\) for each excitatory and inhibitory neuron are formulated as follows:
\begin{widetext}
\begin{eqnarray}
	\tau_{m}\dot{v}_{j}^{(e)}&=&(v_{j}^{(e)})^2+I^{(e)}+2\tau_{m}\left[g^{(ee)}\sum_{l|t_{l}^{(n)}<t}\epsilon_{jl}^{(ee)}\delta(t-t_{l}^{(n)})-g^{(ei)}\sum_{k|t_{k}^{(m)}<t}\epsilon_{jk}^{(ei)}\delta(t-t_{k}^{(m)})\right], \nonumber\\ 
	\tau_{m}\dot{v}_{j}^{(i)}&=&(v_{j}^{(i)})^2+I^{(i)}+2\tau_{m}\left[g^{(ie)}\sum_{l|t_{l}^{(n)}<t}\epsilon_{jl}^{(ie)}\delta(t-t_{l}^{(n)})-g^{(ii)}\sum_{k|t_{k}^{(m)}<t}\epsilon_{jk}^{(ii)}\delta(t-t_{k}^{(m)})\right],
	\label{QIFModel}
\end{eqnarray}
\end{widetext}
where $\tau_m = 30$ ms denotes the membrane time constant, which is consistent across both excitatory and inhibitory neuronal populations. The variables $I^{(e)}$ and $I^{(i)}$ represent the external direct current (DC) applied to the respective neuron populations. The synaptic coupling strength between post-synaptic neurons in population $\alpha$ and pre-synaptic neurons in population $\beta$ is indicated by $g^{(\alpha\beta)}$, where $\alpha$ and $\beta$ can be either $e$ (excitatory) or $i$ (inhibitory). The elements of the adjacency matrix, $\epsilon_{jk}^{(\alpha\beta)}$, are assigned a value of 1 if a connection exists from a pre-synaptic neuron $k$ in population $\beta$ to a post-synaptic neuron $j$ in population $\alpha$, and 0 otherwise. Additionally, the in-degree of neuron $j$ within population $\beta$, represented as $k^{(\alpha\beta)}_j = \sum_{k}\epsilon_{jk}^{(\alpha\beta)}$, quantifies the number of pre-synaptic neurons in population $\beta$ that are connected to neuron $j$ in population $\alpha$. The variable $t_l^{(n)}$ denotes the time of the $n$-th spike emitted by neuron $l$ in population $\alpha$, occurring when the membrane potential is set to $v_{l}^{(\alpha)}(t_{l}^{(n)^{-}}) \rightarrow \infty$ for the spike event. Following each spike emission, the membrane potential is reset to $v_{l}^{(\alpha)}(t_{l}^{(n)^{+}}) \rightarrow -\infty$. It is assumed that the postsynaptic potentials are modeled as $\delta$-pulses and that synaptic transmission occurs instantaneously. 

In order to obtain an effective mean field description for the macroscopic dynamics \cite{diVolo:2018,Bi:2020}, we consider the neurons in both the excitatory and inhibitory populations as being randomly connected. The in-degrees denoted as $k^{(\alpha\alpha)}$, are assumed to follow a Lorentzian distribution:
\begin{equation} 
p\left(k^{(\alpha\alpha)}\right)=\frac{\Delta_{k}^{(\alpha\alpha)}}{\left(k^{(\alpha\alpha)}-K^{(\alpha\alpha)}\right)^2+(\Delta_{k}^{(\alpha\alpha)})^{2}}\cdot \frac{1}{\pi},
\label{Network}
\end{equation}
peaked at $K^{(\alpha \alpha)}$ and with a half-width-half-maximum (HWHM) $\Delta_k^{(\alpha\alpha)}$ measuring the structural heterogeneity in each population. For simplicity, we define $K^{(ee)} = K^{(ii)} = K $. Additionally, we assume that neurons within a population $ \alpha $ are randomly connected to neurons in a distinct population $ \beta \neq \alpha $. In this context, we do not account for structural heterogeneity, maintaining in-degrees at a constant value of $ K^{(ei)} = K^{(ie)} = K $. Tests have demonstrated that employing an Erd\"os-Renyi distribution for the in-degrees $ K^{(ei)} $ and $ K^{(ie)} $, with an average of $ K $, does not influence the observed dynamical behavior \cite{Bi:2021}.

The DC current and synaptic coupling are rescaled using the median in degree, expressed as $ I^{(\alpha)} = \sqrt{K} I_{0}^{(\alpha)} $ and $ g^{(\alpha\beta)} = g_{0}^{(\alpha\beta)}/\sqrt{K} $, to achieve a self-sustained balanced dynamic for the condition where $ N \gg K \gg 1 $ \cite{van:1996,renart:2010,litwin:2012,kadmon:2015}. The parameters reflecting structural heterogeneity are similarly rescaled as $ \Delta_{k}^{(\alpha\alpha)} = \Delta_{0}^{(\alpha\alpha)} \sqrt{K} $, drawing an analogy to Erd\"os-Renyi networks. To reduce the parameter space and meet the requirements necessary for the existence of a balanced state \cite{monteforte:2010}, we establish fixed values for the direct currents, specifically $ I^{(i)}_0 = I^{(e)}_0/1.02 $ with $ I^{(e)}_0 = 0.01 $, the network parameter as $ \Delta^{(ii)}_0 = 0.3 $, and the synaptic couplings as $ g^{(ee)}_0 = 0.27 $, $ g^{(ii)}_0 = 0.953939 $, $ g^{(ie)}_0 = 0.3 $, and $ g^{(ei)}_0 = 0.96286 $ \cite{monteforte:2010}. The subsequent analysis will concentrate on two control parameters, $ K $ and $ \Delta^{(ee)}_0 $, to explore the dynamical transitions between states characterized by different rhythms.

Two indicators are introduced to characterize the macroscopic behavior of such neuronal network. One is the mean membrane potential defined as
\begin{equation}
V^{(\alpha)}(t)=\frac{1}{N^{(\alpha)}}\sum_{i=1}^{N^{(\alpha)}}v^{(\alpha)}_{i}(t).
\label{MeanPotential}
\end{equation}
The other one is the population firing rate, defined as
\begin{equation}
		R^{(\alpha)}(t)=\lim_{\tau_s\to0}\frac{1}{N^{(\alpha)}}\frac{1}{\tau_s}\sum_{j=1}^{N^{(\alpha)}}\sum_{k}\int_{t-\tau_s}^{t}dt^{'}\delta(t^{'}-t_{j}^{k}),
		\label{FiringRate}
\end{equation}
where $t_{j}^{k}$ is the time of the $k$th spike of $j$th neuron, $\delta(t)$ is the Dirac delta function, and we set $\tau_s=10^{-2} \tau_m$.

\section{Next Generation Neural Mass Model}
The intermittent transitions are supposed to be associate with critical behavior. To identify potential occurrences of this behavior, we conduct a bifurcation analysis to locate the critical point. In the thermodynamic limit as $ N \rightarrow \infty $, the macroscopic dynamics of networked QIF neurons characterized by Lorentzian distributed heterogeneity can be effectively represented by a low-dimensional mean-field framework known as the neural mass model \cite{montbrio:2015}. Within this neural mass model, the dynamics of each neuronal population are described using two primary variables: the mean membrane potential $ V^{(\alpha)}(t) $ and the instantaneous firing rate $ R^{(\alpha)}(t) $. The following section outlines the main steps involved in deriving this mean-field formulation specifically for sparsely coupled excitatory and inhibitory networks of QIF neurons.

At first, we introduce the non-dimensional time as $\tilde{t}=\frac{t}{\tau_m}$. The other variables are expressed as $r^{(\alpha)}=\tau_m R^{(\alpha)}$, $v^{(\alpha)}=V^{(\alpha)}$, $q^{(\alpha)}_2=Q^{(\alpha)}_2$, and $p^{(\alpha)}_2 = P^{(\alpha)}_2$. The quenched disorder related to the in-degree distribution can be articulated in terms of random synaptic couplings. Specifically, each neuron $ i $ within population $ \alpha $ is subjected to currents characterized by the amplitude $ g^{(\alpha\beta)}_{0}k_{i}^{(\alpha\beta)}r^{(\beta)}/(\sqrt{K}) $, which is directly proportional to their in-degrees $ k_{i}^{(\alpha\beta)} $, where $ \beta $ can be either excitatory ($ e $) or inhibitory ($ i $). Consequently, we can conceptualize the neurons as being fully interconnected, albeit with random coupling values that follow a Lorentzian distribution, with a median of $ g^{(\alpha\beta)}_{0}\sqrt{K} $ and a half-width at half-maximum (HWHM) of $ g^{(\alpha\beta)}_{0}\Delta^{(\alpha\beta)}_{0} $.

In full generality, we can assume that the synaptic coupling $g_{i}^{(\alpha)}$ for neuron $i$ follows a Lorentzian distribution $h(g^{(\alpha)})$, characterized by a median $g^{(\alpha\beta)}_{0}\sqrt{K}$ and HWHM $g^{(\alpha\beta)}_{0}\Delta^{(\alpha\beta)}_{0}$. Within the thermodynamic limit, the dynamics of the network, as described by Equation (\ref{QIFModel}), can be analyzed through the probability density function (PDF) $\rho(v^{(\alpha)}, t|g^{(\alpha)})$, which satisfies the corresponding Fokker–Planck equation (FPE):
\begin{eqnarray} \nonumber
	&\partial_{t}p(v^{(\alpha)},t|g^{(\alpha)})+\partial_{v^{(\alpha)}}\left[((v^{(\alpha)})^{2}+I_{g}^{(\alpha)})p(v^{(\alpha)},t|g^{(\alpha)})\right] \\
	&=\sigma^{2}\partial_{v^{(\alpha)}}^{2}p(v^{(\alpha)},t|g^{(\alpha)}),
	\label{FPE}
\end{eqnarray}
where $I_{g^{(\alpha)}}^{(\alpha)}=g^{(\alpha)}r^{(\alpha)}(t)$. With the absence of noise the solution of Eq. (\ref{FPE}) converges to a Lorentzian distribution for any initial PDF $p(v^{(\alpha)},0|g^{(\alpha)})$:
\begin{equation}
	p(v^{(\alpha)},t|g^{(\alpha)})=\frac{a^{(\alpha)}_{g}}{[\pi((a^{(\alpha)}_{g})^{2}+(v^{(\alpha)}-v^{(\alpha)}_{g^{(\alpha)}})^{2})]},
	\label{model}
\end{equation}
where $v^{(\alpha)}_g$ is the mean membrane potential and
\begin{equation}
	r^{(\alpha)}_{g^{(\alpha)}}(t)=\lim_{v\rightarrow\infty}(v^{(\alpha)})^{2}p(v^{(\alpha)},t|g^{(\alpha)})=\frac{a^{(\alpha)}_{g^{(\alpha)}}}{\pi}
	\label{Vg}
\end{equation}
is the firing rate for the $g^{(\alpha)}$-subpopulation. 

In the case of the sparse deterministic systems under the Poissonian approximation for the input spike trains, one can introduce the characteristic function for $v_g$, which is the Fourier transform of its PDF:
\begin{equation}
	\mathcal{F}_{g^{(\alpha)}}(k)=\langle e^{ikv^{(\alpha)}_{g^{(\alpha)}}} \rangle = P. V. \int_{-\infty}^{\infty}e^{ikv^{(\alpha)}_{g^{(\alpha)}}}p(v^{(\alpha)},t|g^{(\alpha)})\d v^{(\alpha)}_{g^{(\alpha)}},
	\label{Fourier}
\end{equation}
where $P. V.$ indicates the Cauchy principal value. In this framework the FPE Eq. (\ref{FPE}) can be rewritten as
\begin{equation}
	\partial_t F_{g^{(\alpha)}}= ik[I_{g^{(\alpha)}}\mathcal{F}_{g^{(\alpha)}}-\partial_{k}^{2}\mathcal{F}_{g^{(\alpha)}}]-\sigma^{2}k^{2}\mathcal{F}_{g^{(\alpha)}}.
	\label{FPF2}
\end{equation}

Under the assumption that $\mathcal{F}_{g^{(\alpha)}}(k, t)$ is an analytic function of  $g^{(\alpha)}$, one can estimate the characteristic function averaged over the heterogeneous population:
\begin{equation}
	F(k,t)=\int dg^{(\alpha)} \mathcal{F}_{g^{(\alpha)}}(k,t)h(g^{(\alpha)}),
	\label{FPF3}
\end{equation}
and via the residue theorem the corresponding FPE yields
\begin{equation}
	\partial_t F= ik[H_0 F-\partial_{k}^{2}F]-|k|D_{0}F-\sigma^{2}k^{2}F,
	\label{FPF4}
\end{equation}
where $H_0=g^{(\alpha\alpha)}_{0}\sqrt{K} r^{(\alpha)}$ and $D_0= g^{(\alpha\alpha)}_{0}\Delta^{(\alpha\alpha)}_{0} r^{(\alpha)}$.

With the logarithm of the characteristic function, $\Phi(k,t)=\ln (F(k,t))$, one obtains the evolution equation
\begin{equation}
	\partial_{t}\Phi=ik[H_0-\partial_{k}^{2}\Phi-(\partial_k \Phi)^2]-|k|D_0-\sigma^{2}k^{2}.
	\label{FPF5}
\end{equation}

In this context the Lorentzian ansatz amounts to $\Phi_L=ikv^{(\alpha)}-a^{(\alpha)}|k|$. By substituting $\Phi_L$ into Eq. (\ref{FPF4}) for $\sigma=0$ one gets 
\begin{equation}
	\dot{v}^{(\alpha)}=H_0 +(v^{(\alpha)})^2 - (a^{(\alpha)})^2, \  \  \dot{a}^{(\alpha)}=2a^{(\alpha)}v^{(\alpha)}+D_0,
	\label{FPF6}
\end{equation}
which coincides with the two dimensional exact mean-field dynamics reported in Ref. \cite{montbrio:2015} with $r^{(\alpha)}=a^{(\alpha)}/\pi$.

For a finite-size system, in order to consider deviations from the case of Lorentzian distribution the following general polynomial form for $\Phi$ is introduced \cite{goldobin:2021}:
\begin{equation}
	\Phi = -a^{(\alpha)}|k| +ikv^{(\alpha)} - \sum_{n=2}^{\infty}\frac{q^{(\alpha)}_n |k|^n +ip^{(\alpha)}_n |k|^{n-1}k}{n}.
	\label{FPF7}
\end{equation}

A set of complex pseudo-cumulants are also proposed:
\begin{equation}
	W_1=a^{(\alpha)}-iv^{(\alpha)}, \  \  W_n=q^{(\alpha)}_n+ip^{(\alpha)}_n.
	\label{FPF8}
\end{equation}

By inserting the expansion Eq. (\ref{FPF7}) into the Eq. (\ref{FPF5}) one gets the evolution equations for the pseudo-cumulants:
\begin{align}
	\dot{W}_m&=(D_0-iH_0)\delta_{1m}+2\sigma^{2}\delta_{2m} \nonumber\\ 
	&+im\left(-mW_{m+1}+\sum_{n=1}^{m}W_{n}W_{m+1-n}\right),
	\label{FPF9}
\end{align}
where $\delta_{ij}$ is the Kronecker delta function and for simplicity we assumed $k>0$. It is important to notice that the time-evolution of $W_m$ depends only on the pseudo-cumulants up to the order $m+1$, therefore, the hierarchy of equations can be easily truncated at the $m$-th order by setting $W_{m+1} = 0$. As shown in Ref. \cite{goldobin:2021} the modulus of the pseudo-cumulants exhibits a scaling relationship of the form $|W_m| \propto \sigma^{2(m-1)}$ with the noise amplitude $\sigma$. This scaling justifies the consideration of an expansion confined to the initial few orders of pseudo-cumulants.

In this work, we consider the pseudo-cumulants Eq. (\ref{FPF8}) up to the second order to obtain the corresponding neural mass model, which leads to the following mean-field formulation describing the low-dimensional behavior of Eq. (\ref{QIFModel}):
	\begin{eqnarray}
		\dot{r}^{(\alpha)}&=&2r^{(\alpha)}v^{(\alpha)}+(\Delta^{(\alpha\alpha)}_0|g^{(\alpha\alpha)}_{0}|r^{(\alpha)}+p_{2}^{(\alpha)})\pi^{-1}, \nonumber\\
		\dot{v}^{(\alpha)}&=&(v^{(\alpha)})^{2}-(\pi r^{(\alpha)})^{2}+\sqrt{K}(I^{(\alpha)}  \nonumber\\ 
		&+&g^{(\alpha\alpha)}_{0} r^{(\alpha)}+g^{(\alpha\beta)}_{0} r^{(\beta)})+ q_{2}^{(\alpha)},\nonumber\\
		\dot{q}_{2}^{(\alpha)}&=&2\mathcal{N_{R}}+4(p_3^{(\alpha)}+q_2^{(\alpha)}v^{(\alpha)}-\pi p_2^{(\alpha)} r^{(\alpha)}),\nonumber\\
		\dot{p}_{2}^{(\alpha)}&=&2\mathcal{N_{I}}+4(-q_3^{(\alpha)}+p_2^{(\alpha)}v^{(\alpha)}+\pi q_2^{(\alpha)} r^{(\alpha)}),
		\label{Massnon}
	\end{eqnarray}
	with $\mathcal{N_R}=\frac{(g^{(\alpha\alpha)}_{0})^2 r^{(\alpha)}}{2K}+\frac{(g^{(\alpha\beta)}_{0})^{2}r^{(\beta)}}{2K}$,  $\mathcal{N_I}=-\frac{(g^{(\alpha\alpha)}_{0})^2 \Delta^{(\alpha\alpha)}_0 r^{(\alpha)}}{2K}=-\Delta^{(\alpha\alpha)}_0 \mathcal{N_R}$, and $\Delta_0^{(\alpha\beta)}=\Delta_0^{(\beta\alpha)}=0$. Due to the assumption that the connections among neurons of different populations are random but with a fixed in-degree $K^{(\alpha\beta)}=K^{(\beta\alpha)}=K$.  The macroscopic variables $r^{(\alpha)}$ and $v^{(\alpha)}$ represent the population firing rate and the mean membrane potential, while the terms $q_2^{(\alpha)}, p_2^{(\alpha)}$ take into account the dynamical modification of the PDF of the membrane potentials with respect to a Lorentzian profile. In addition, we set  $q_3^{(\alpha)}=0$, $p_3^{(\alpha)}=0$.
	
	Finally, with the time scale $\tau_m$ back, the macroscopic dynamics of the population of QIF neurons (1) is exactly described as follows:
		\begin{eqnarray}
			\tau_m\dot{R}^{(\alpha)}&=&2R^{(\alpha)}V^{(\alpha)}+(\Delta^{(\alpha\alpha)}_0|g^{(\alpha\alpha)}_{0}|R^{(\alpha)}+P_{2}^{(\alpha)}/\tau_m)\pi^{-1}, \nonumber\\
			\tau_m\dot{V}^{(\alpha)}&=&(V^{(\alpha)})^{2}-(\pi \tau_m R^{(\alpha)})^{2}+\sqrt{K}(I^{(\alpha)}  \nonumber\\ 
			&+&g^{(\alpha\alpha)}_{0}\tau_m R^{(\alpha)}+g^{(\alpha\beta)}_{0}\tau_m R^{(\beta)})+ Q_{2}^{(\alpha)},\nonumber\\
			\tau_m\dot{Q}_{2}^{(\alpha)}&=&2\mathcal{\tilde{N}_{R}}+4(P_3^{(\alpha)}+Q_2^{(\alpha)}V^{(\alpha)}-\pi Q_2^{(\alpha)} \tau_m R^{(\alpha)}),\nonumber\\
			\tau_m\dot{P}_{2}^{(\alpha)}&=&2\mathcal{\tilde{N}_{I}}+4(-Q_3^{(\alpha)}+P_2^{(\alpha)}V^{(\alpha)}+\pi P_2^{(\alpha)}\tau_m R^{(\alpha)}),
			\label{Masstime}
		\end{eqnarray}
		with $\mathcal{\tilde{N}_R}=\tau_m\mathcal{N_R}$,  $\mathcal{\tilde{N}_I}=\tau_m\mathcal{N_I}=-\Delta^{(\alpha\alpha)}_0 \tau_m\mathcal{N_R}$.

\begin{figure}
	\includegraphics[width=1.0\linewidth]{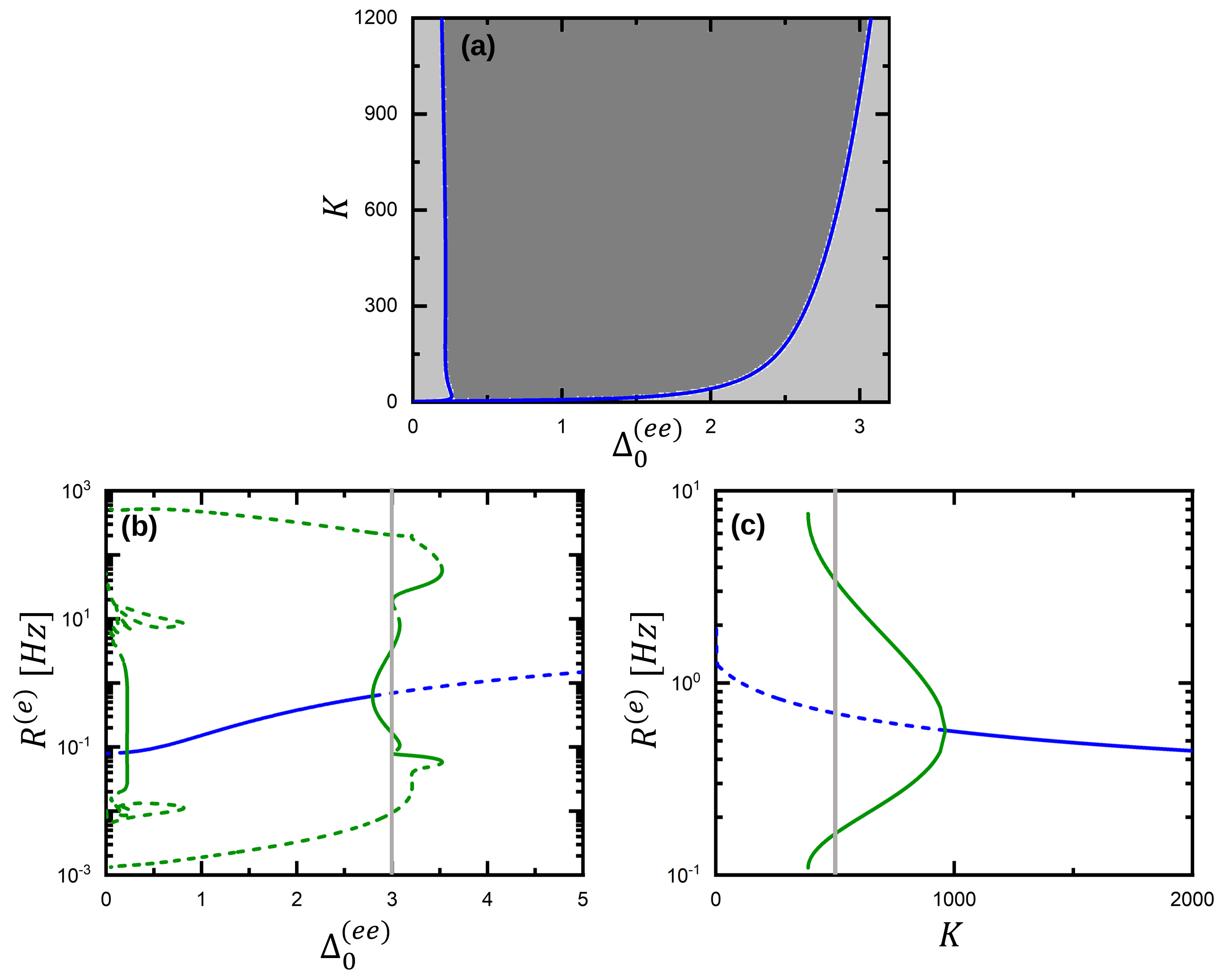} \caption{(color online). (a) The phase diagram obtained by Eq. (\ref{Masstime}). The dark gray area in the middle indicates a state with a stable focus and unstable limit cycles, the light gray areas for large and small $\Delta^{(ee)}_0$ indicate a state with stable limit cycles and a unstable focus. (b) Bifurcation diagram of firing rate $R^{(e)}$ depending on $\Delta^{(ee)}_0$ for Eq. (\ref{Masstime}) with $K=500$.  (c) Bifurcation diagram of firing rate $R^{(e)}$ depending on $K$ with $\Delta^{(ee)}_0=3$.  The  solid (dashed) blue  line indicates the stable (unstable) focus. The solid (dashed) green lines represent the stable (unstable) limit cycles. The grays line mark  the parameters of $\Delta^{(ee)}_0=3$ and $K=500$ where we start the investigation of the transition behavior between different frequency bands. 			  }
	\label{bifurcation}
\end{figure}

Given that excitatory neurons significantly outnumber inhibitory neurons, the overall dynamics of the system is predominantly dependent on the behavior of the excitatory population. This assertion has been verified through comparisons with empirical data \cite{martinez:2021}. Consequently, in the subsequent bifurcation analysis of Equation (\ref{Masstime}), we will concentrate on the dynamics of the variables associated with the excitatory population.

Figure \ref{bifurcation}(a) illustrates the phase diagram in the $\Delta^{(ee)}_0$ versus $K$ parameter space as derived from Equation (\ref{Masstime}). Within the intermediate range of $\Delta^{(ee)}_0$, a stable focus is observed coexisting with unstable limit cycles. Conversely, for both large and small values of $\Delta^{(ee)}_0$, the focus becomes unstable, giving rise to stable limit cycles, which signifies the presence of oscillatory dynamics. Therefore, to observe an oscillating average membrane potential and potential critical behaviors, we begin our analysis with parameters at the limit cycle and close to the Hopf bifurcation point, specifically $\Delta^{(ee)}_0=3$ and $K=500$. Figures \ref{bifurcation}(b) and (c) further illustrate the bifurcation diagrams for the system. With $\Delta^{(ee)}_0=3$ and $K=500$ (marked by the gray lines), there existing a stable limit cycle (green solid lines), an unstable focus (blue dashed line) and an unstable limit cycle (green dashed lines), confirming the presence of an oscillating $V$. From \ref{bifurcation}(b) we observe that the selected parameters are situated near the bifurcation point, indicating that the system is close to criticality, where transitions between different states may occur.

\section{Intermittent transitions between $\delta$ and $\theta$ bursts} 

\begin{figure}
	\includegraphics[width=0.7\linewidth]{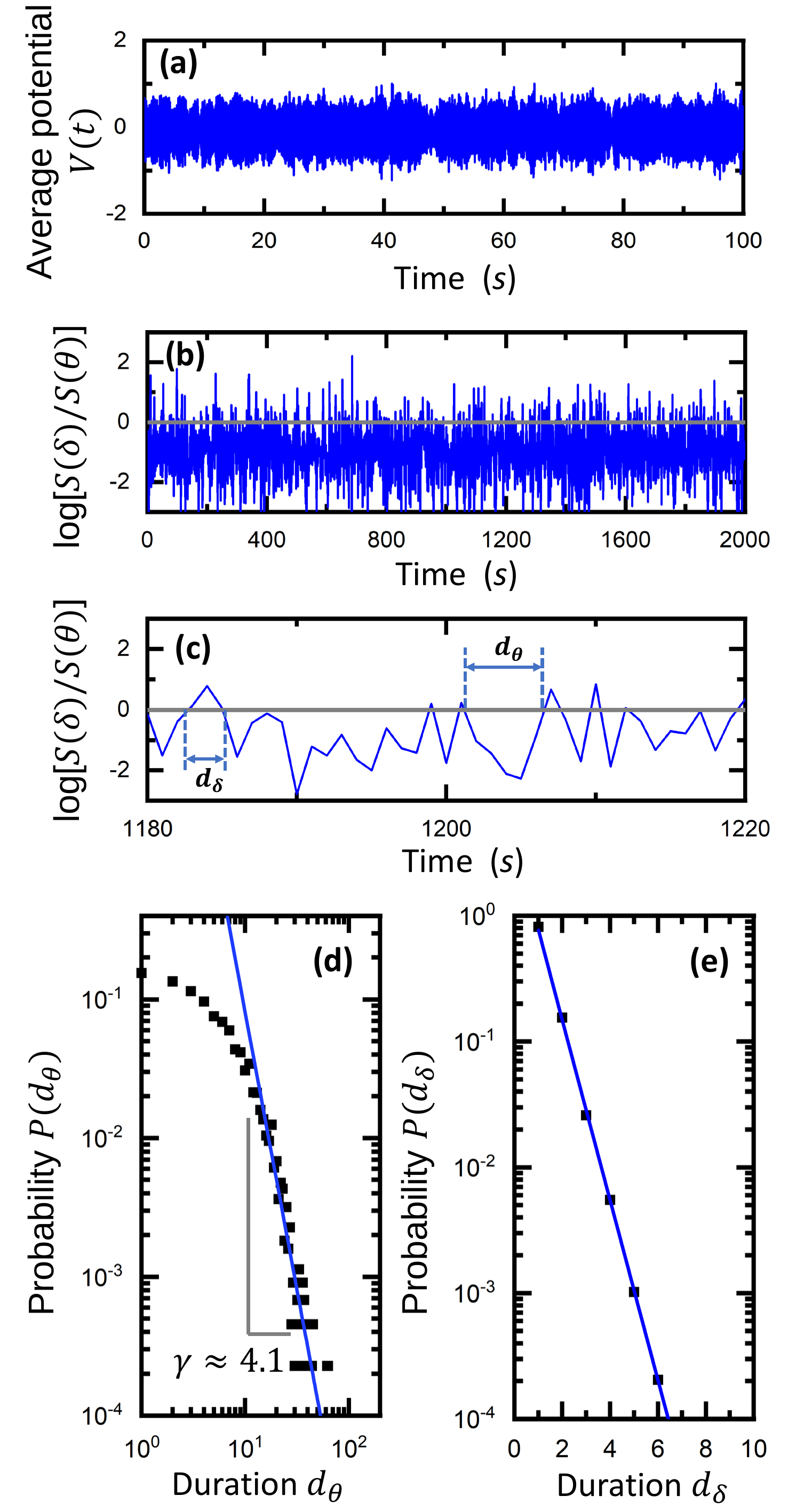} \caption{(color online). (a) Time series of the mean membrane potential generated by the networked QIF model Eq. (\ref{QIFModel}). (b) The time-varying radio $R_{\delta\theta}=S(\delta)/S(\theta)$ of the spectral power in $\delta$ (0 - 4 Hz) and $\theta$ (4 - 8 Hz) bands obtained for each time window of $1$ s. The vertical value is plotted in logarithmic scale. Therefore, the $log(R_{\delta\theta})>0$ (above the gray horizontal line) indicates the predominance of $\delta$ rhythm, while $log(R_{\delta\theta})<0$ (below the gray horizontal line) corresponds to the predominance of $\theta$ rhythm. (c) The duration of states dominated by $\theta$ and $\delta$ bands. (d) Probability density distribution for the duration of states dominated by $\theta$ band ($P(d_{\theta})$) plotted in the  double-logarithmic coordinates (black squares). For the range of $d_\theta>10$ s,  $P(d_{\theta})$ follows a power-law behavior of $P(d_{\theta})\sim d^{-\gamma}_{\theta}$ with a component of $\gamma\approx 4.1$ indicated by the blue solid line. (e) Probability density distribution for the duration of states dominated by $\delta$ band ($P(d_{\delta})$) plotted in the  linear-logarithmic coordinates (black squares). The curve exhibits an exponential behavior indicated by the blue solid line. The parameters are $\Delta^{(ee)}_0=3$, $K=500$, the simulation is performed in a network with size $N^{(e)}=5000$ and $N^{(i)}=1000$. In the simulation we use the time series with $t>60$ s to guarantee that the system has reached stationary. Statistics are then done based on simulated time series of mean membrane potential $V$ with length of 20000 s.  }
	\label{timeseries}
\end{figure}

\begin{figure}
	\includegraphics[width=0.7\linewidth]{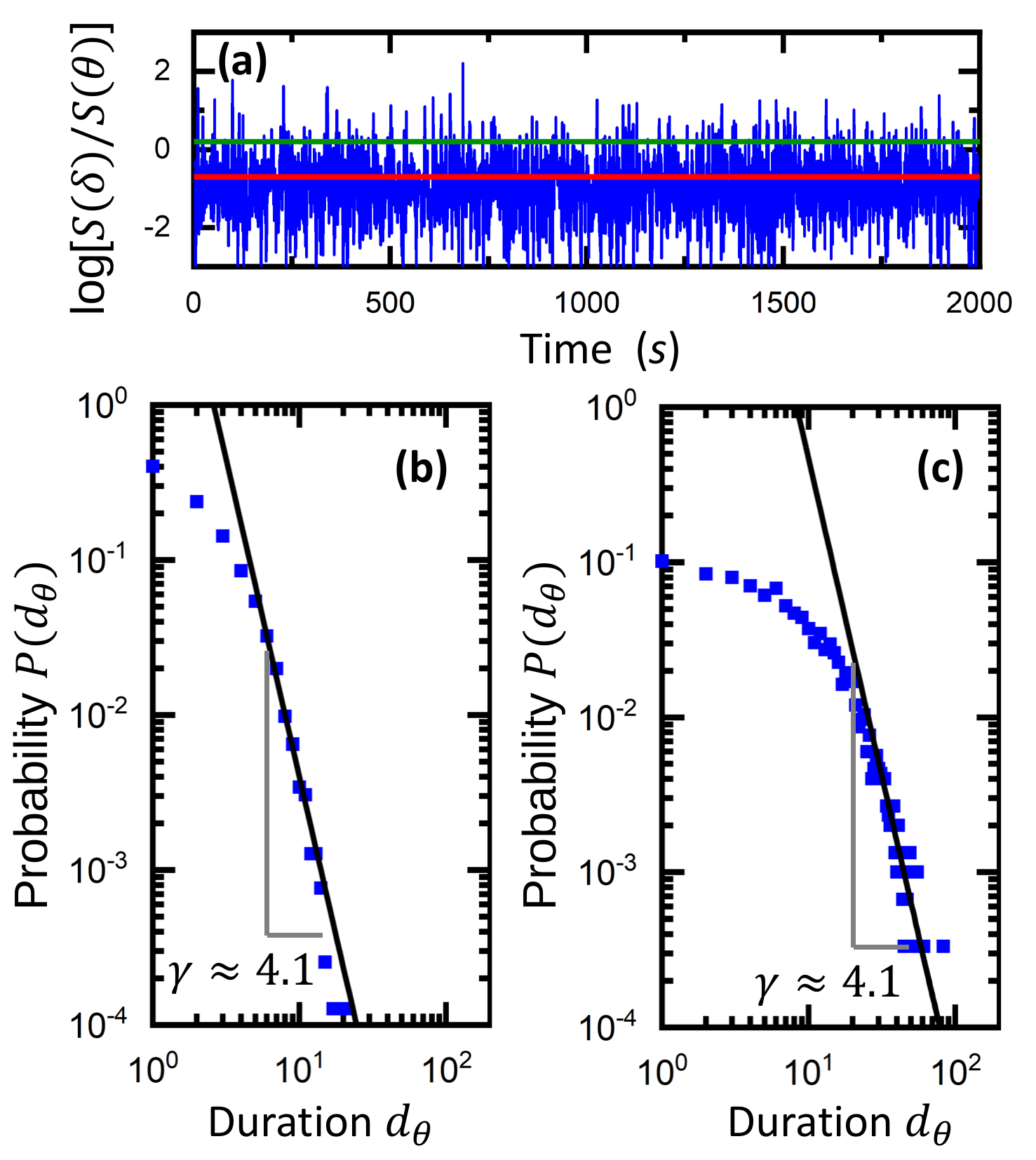} \caption{(color online). (a) Different threshold to determine the predominance of $\delta$ and $\theta$ rhythms with $\log(R_{\delta\theta})=-0.693$ (red line, corresponding to $R_{\delta\theta}=0.5$) and $\log(R_{\delta\theta})=0.182$ (green line, corresponding to $R_{\delta\theta}=1.2$). Probability density distribution for the duration of states dominated by $\theta$ band ($P(d_{\theta})$) are plotted in the  double-logarithmic coordinates (black squares) for the case of (b): $\log(R_{\delta\theta})=-0.693$, and (c): $\log(R_{\delta\theta})=0.182$. The power-law behavior observed for the tails of the distribution dose not change with the threshold determining the predominance of $\delta$ and $\theta$ rhythms. }
	\label{threshold}
\end{figure}

Figure \ref{timeseries}(a) presents the time series of the mean membrane potential as derived from Equation (\ref{QIFModel}) for $K=500$ and $\Delta^{(ee)}_0=3$. This time series is then analyzed by evaluating the spectral power across distinct frequency bands utilizing non-overlapping windows of one second in length ($w=1$ s). Within each window, we compute the spectral power for the $\delta$-wave band (0 - 4 Hz) and the $\theta$-wave band (4 - 8 Hz). To investigate the transition behavior between states characterized by these two rhythms, utilizing the same approach as in the empirical researches \cite{Wang:2019,Huo:2024,Lombardi:2020}, we calculate the ratio $ R_{\delta\theta} = S(\delta)/S(\theta) $, which reflects the relationship between the spectral powers of the $\delta$ and $\theta$ bands. With a threshold of $R_{\delta\theta}=1$ the time series is distincted into two states. In Figure \ref{timeseries}(b) we plot the fluctuations of $\log( R_{\delta\theta})$, similar as in the empirical researches \cite{Wang:2019,Huo:2024,Lombardi:2020}, the time windows with $\log(R_{\delta\theta})>0$ (above the threshold of $R_{\delta\theta}=1$) are noted as $\delta$-bursts, and time windows with $\log(R_{\delta\theta})<0$ are marked as $\theta$-bursts. The fluctuations of $\log( R_{\delta\theta}) $ between values greater than zero and those less than zero indicate intermittent transitions between the $\delta$- and $\theta$-bursts. Then the duration for $\theta$ ($ d_{\theta}$) and $\delta$ ($ d_{\delta}$) bursts are calculated (Figure \ref{timeseries}(c)). The probability density distributions for $ d_{\theta}$ and $ d_{\delta}$, denoted as $ P(d_{\theta}) $ and $ P(d_{\delta}) $, are calculated using following linear binning procedure. Given the window size $w=1$, the bin boundaries $e_1,e_2,\dots,e_n$ are obtained using the recursive relation $e_n=e_1+w\sum_{i=2}^{n}b$, with $e_1=0.5w$, $b=1$ and $n\ge2$. We also checked that a logarithmic binning, i.e. linear binning in logarithmic scale, does not qualitatively change the observed power-law behavior.

In Figures \ref{timeseries}(d) and \ref{timeseries}(e), $ P(d_{\theta}) $ and $ P(d_{\delta}) $ are shown in double-logarithmic and linear-logarithmic plots, respectively. For durations exceeding 10 seconds for the $\theta$-bursts ($ d_{\theta} > 10 $ s, at about the time scale of minute), the distribution $ P(d_{\theta}) $ approximates a power-law behavior: $P(d_{\theta})\sim d^{-\gamma}_{\theta}$, with an exponent of $ \gamma \approx 4.1 $. Conversely, $ P(d_{\delta}) $ demonstrates an exponential distribution. This finding is qualitatively consistent with empirical observations reported by Wang et al. \cite{Wang:2019}. 

The power-law exponent for the duration of $\theta$ bursts, which is a key feature of critical dynamics, remains unaffected by changes in the threshold of $R_{\delta\theta}$ used to differentiate between the $\delta$ and $\theta$ rhythms, as demonstrated in Ref. \cite{Wang:2019}. To further align our findings with empirical data, we tested two different thresholds: $R_{\delta\theta}=0.5$ (red line) and $R_{\delta\theta}=1.2$ (green line), as illustrated in Figure \ref{threshold}(a). We found that the power-law behavior for the duration of $\theta$ bursts is maintained, and the exponent $\gamma \approx 4.1$ remains unchanged (see Figure \ref{threshold}(b) and (c)). This aligns with empirical observations suggesting dynamics near criticality. Additionally, Figure \ref{bifurcation}(b) supports this, showing that the selected parameter is near the Hopf bifurcation point, which is a critical point separating the stable focus and limit cycle.

\section{Effect of intrinsic fluctuation}  
The bifurcation analysis elucidates the potential stationary states of the system in the limit as $ N\rightarrow \infty$, under the assumption of no perturbations or fluctuations. In this scenario, a stable limit cycle is identified as a stationary state characterized by oscillations of $ V^{(e)} $ at a constant frequency. To adequately address the observed intermittent behavior, it is essential to incorporate additional mechanisms, such as the effects of intrinsic fluctuations that are characteristic of finite-size systems \cite{Daido:1990,Volo:2022}. 

\begin{figure}
	\includegraphics[width=0.9\linewidth]{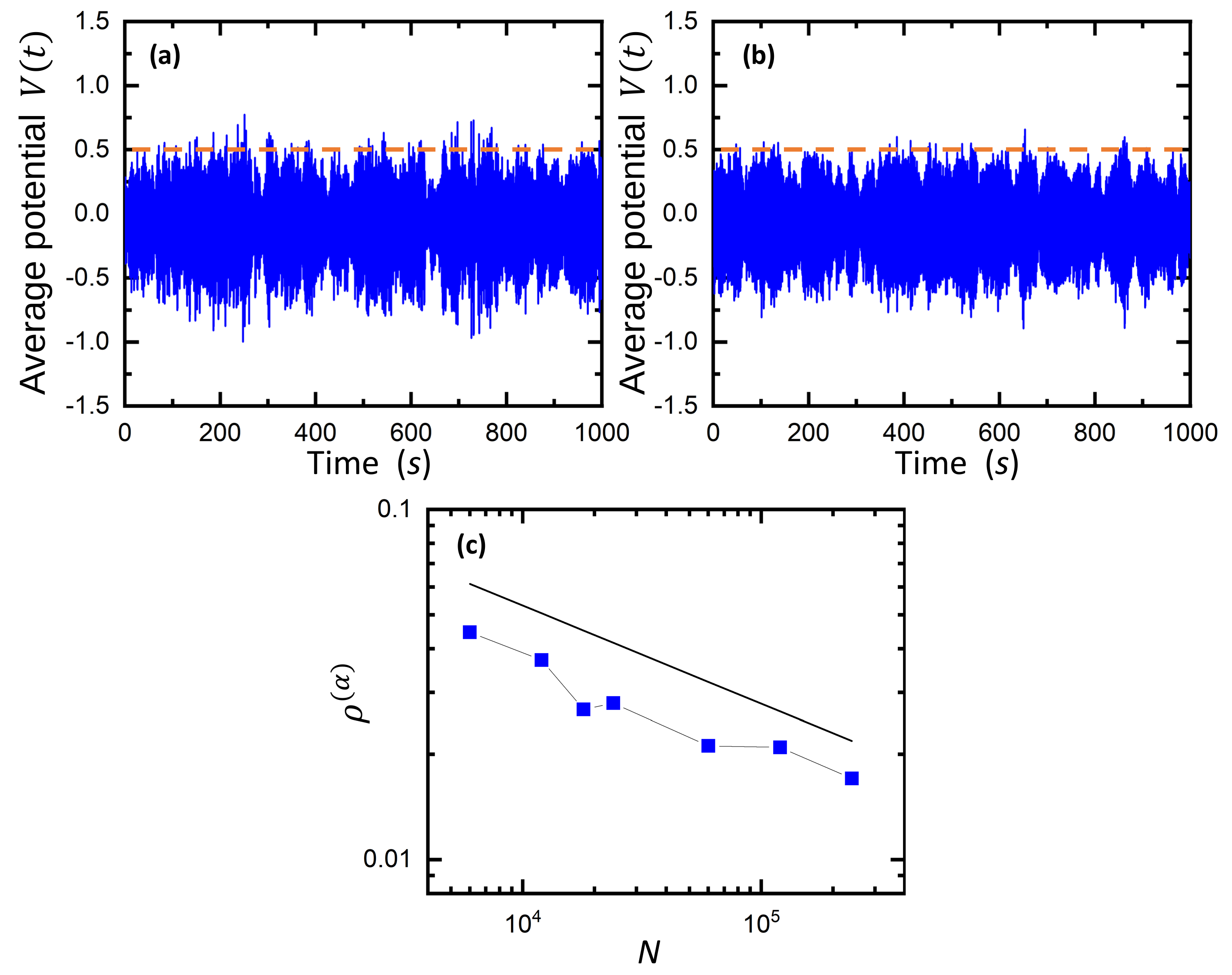} \caption{(color online). Time series of the mean membrane potential generated by the networked QIF model Eq. (\ref{QIFModel}) for (a): $N=6000$ ($N^{(e)}=5000$ and (b): $N^{(i)}=1000$) and $N=18000$ ($N^{(e)}=15000$ and $N^{(i)}=3000$). Orange dashed line marks the value of 0.5. One can see that there are more peaks high than 0.5 in (a) than in (b).  Parameters are same as in Figure \ref{timeseries}. (c) $\rho^{(\alpha)}$, the level of coherence in the neural activity, decreases with the system size N following a scaling behavior $\rho^{(\alpha)}\sim N^{-a}$.}
	\label{size}
\end{figure}

To exemplify this fluctuation behavior, we present the average membrane potential derived from Equation (\ref{QIFModel}) for $ N=6000 $ (Figure \ref{size} (a)) and $ N=18000 $ (Figure \ref{size} (b)). It is evident that there are more peaks exceeding 0.5 (indicated by the orange dashed line) for $ N=6000 $ compared to $ N=18000 $. This observation suggests that the fluctuations in $ V $ are more pronounced in smaller systems. To quantitatively assess how intrinsic fluctuations vary with system size, we compute $ \rho^{(\alpha)} $, which represents the level of coherence in neural activity for population $\alpha$ defined as
\begin{equation}
	\rho^{(\alpha)}=\left(\frac{\sigma_{v^{(\alpha)}}^2}{\sum_{i=1}^{N^{(\alpha)}}\sigma_i^2/N^{(\alpha)}} \right)^{1/2},
	\label{rho}
\end{equation}
where $\sigma_{V^{(\alpha)}}$ is the standard deviation of the mean membrane potential,  $\sigma_i=\left\langle (v^{(\alpha)})_i^2\right\rangle-\left\langle (v^{(\alpha)})_i\right\rangle^2$ and $\left\langle \cdot\right\rangle$ denotes a time average. For an infinite size system undergoing  asynchronous dynamics (focus or limit cycle near the bifurcation point), the parameter $\rho^{(\alpha)}$ approaches zero. Conversely, in finite-sized systems, fluctuations can result in a $\rho^{(\alpha)}$ value that is greater than zero. As the size of the system increases, these fluctuations diminish, leading to a decrease in $\rho^{(\alpha)}$ that follows the relationship $\rho^{(\alpha)} \sim N^{-a}$, as illustrated in Figure \ref{size} (c). This observation substantiates the presence of intrinsic fluctuations that are characteristic of finite-sized systems \cite{Volo:2022}. Such intrinsic fluctuations are an inevitable aspect of finite-sized systems and can facilitate transitions among multiple stable stationary states \cite{goldobin:2021,renart:2007,Klinshov:2024}, and influencing the dynamics of the system \cite{Pyragas:2023,Klinshov:2022,Goldobin:2024}.  

\begin{figure}
	\includegraphics[width=0.9\linewidth]{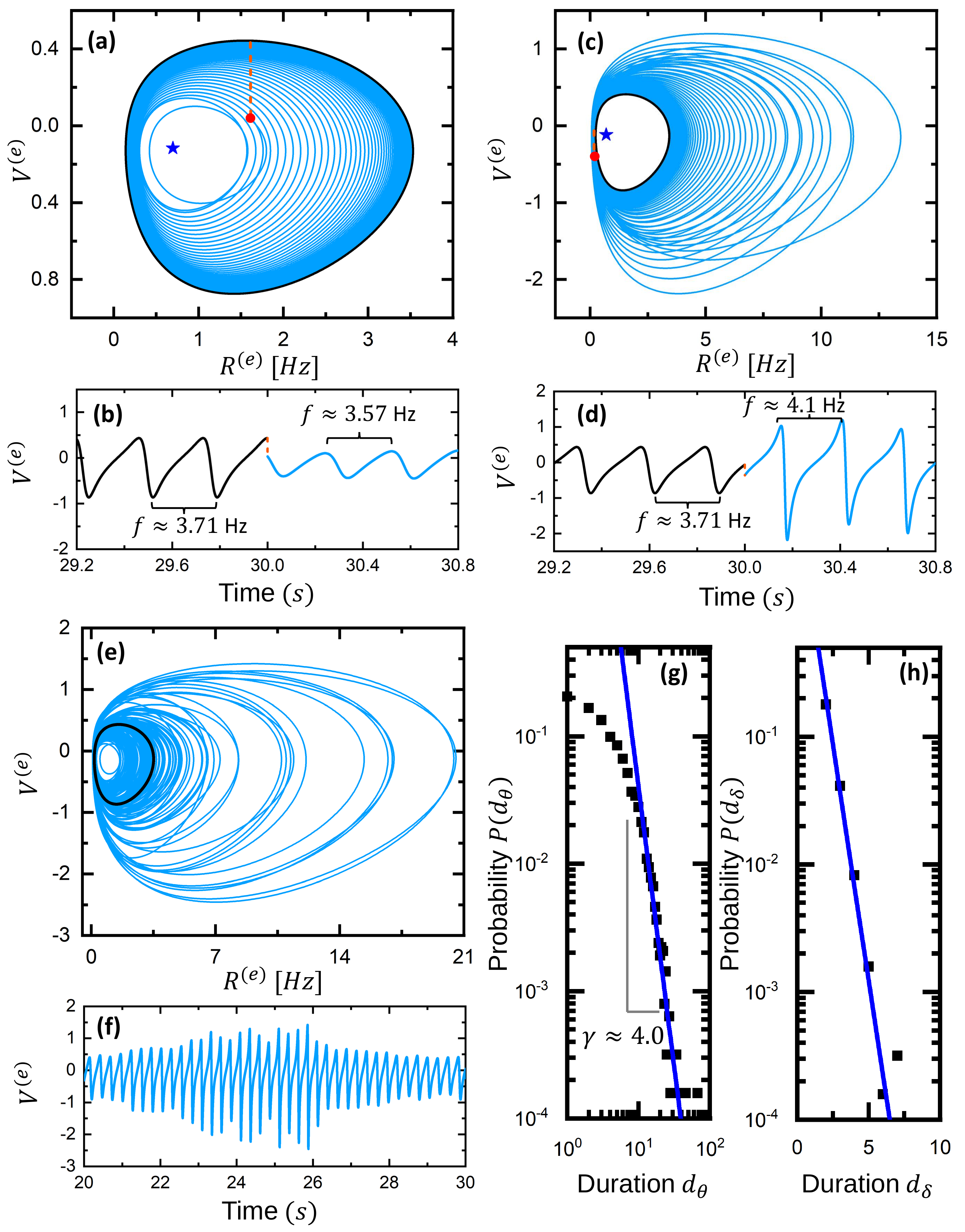} \caption{(color online). Effect of the intrinsic fluctuation to the dynamics of the system. In the case of $\Delta^{(ee)}_0=3$ and $K=500$, the same as in Fig. \ref{timeseries}, there are a stable limit cycle (black line), an unstable focus (blue star) and an unstable limit cycle outside the stable one (not shown). Each of them corresponds to a particular oscillation frequency.  (a) The dynamical behavior of the system follows the stable limit cycle (black line) if there is no fluctuation. After a perturbation kicks the system towards the unstable focus (marked by the red dot and dashed line), the system will start to move back towards the stable limit cycle with rotation (blue line). (b) The oscillation frequency of the mean membrane potential $V^{(e)}$  changes before and after the perturbation. (c) If the perturbation kicks the system towards the unstable limit cycle (marked by the red dot and dashed line), the system will also move back to the stable limit cycle with rotation (marked by the red dot and dashed line). (d) Correspondingly, $V^{(e)}$ also changes before and after the perturbation. (e) When there is a continuous perturbation or intrinsic fluctuation the system cannot stay fixed on a particular state, it will rotate around the stable limit cycle, i.e., sometimes close to the unstable limit cycle and sometimes towards the unstable focus. (f) As a result, we observe constantly changing amplitude and oscillation frequency for $V^{(e)}$. (g) and (h): Statistical analysis on the time series of $V^{(e)}$ in (f),  same behavior as for the QIF model Eq. (\ref{QIFModel}) shown in Fig. \ref{timeseries} are observed.	The continuous perturbation is added for $\dot{V}^{(e)}$ and $\dot{V}^{(i)}$ as additive uniform white noise with zero mean in the range $[-0.0005,0.0005]$.	  }
	\label{perturbation}
\end{figure}

Since the low-dimensional equation (\ref{Masstime}) describes the dynamics for a system with infinite size. Therefore, as demonstrated in Figure \ref{perturbation}(a) for $K=500$ and $\Delta^{(ee)}_0=3$, in the absence of perturbations or fluctuations, the dynamics of the system are restricted to a stable limit cycle characterized by a constant rotational frequency. However, the introduction of a perturbation can cause the system to deviate from this stable state (red dot in Figs. \ref{perturbation}(a) and (c)). Subsequently, the system initiates a return to stability through rotation, as the neighboring stationary states are unstable, as illustrated by the blue trajectory in Figures \ref{perturbation}(a) and (c). When the perturbation is directed towards the unstable focus inside, we observe an oscillation of $V^{(e)}$ characterized by a reduced amplitude and a slower frequency (as shown in Fig. \ref{perturbation}(a) and (b)). Conversely, if the perturbation is directed towards the unstable limit cycle outside, both the oscillation amplitude and frequency of $V^{(e)}$ increase (as demonstrated in Fig. \ref{perturbation}(c) and (d)).

In finite-size systems, intrinsic fluctuations induce continuous perturbations that prevent the system from maintaining on the stable limit cycle; instead, the system's rotation vibrates around this limit cycle. To incorporate the influence of fluctuations into the mass model, for simplicity, we introduce a noise term into Eq. (\ref{Masstime}):
\begin{eqnarray}
	\tau_m\dot{R}^{(\alpha)}&=&2R^{(\alpha)}V^{(\alpha)}+(\Delta^{(\alpha\alpha)}_0|g^{(\alpha\alpha)}_{0}|R^{(\alpha)}+P_{2}^{(\alpha)}/\tau_m)\pi^{-1}, \nonumber\\
\tau_m\dot{V}^{(\alpha)}&=&(V^{(\alpha)})^{2}-(\pi \tau_m R^{(\alpha)})^{2}+\sqrt{K}(I^{(\alpha)}  \nonumber\\ 
&+&g^{(\alpha\alpha)}_{0}\tau_m R^{(\alpha)}+g^{(\alpha\beta)}_{0}\tau_m R^{(\beta)})+ Q_{2}^{(\alpha)}+\xi^{(\alpha)}(t),\nonumber\\
\tau_m\dot{Q}_{2}^{(\alpha)}&=&2\mathcal{\tilde{N}_{R}}+4(P_3^{(\alpha)}+Q_2^{(\alpha)}V^{(\alpha)}-\pi P_2^{(\alpha)} \tau_m R^{(\alpha)}),\nonumber\\
\tau_m\dot{P}_{2}^{(\alpha)}&=&2\mathcal{\tilde{N}_{I}}+4(-Q_3^{(\alpha)}+P_2^{(\alpha)}V^{(\alpha)}+\pi Q_2^{(\alpha)}\tau_m R^{(\alpha)}),
	\label{NoiseMass}
\end{eqnarray}
where $\xi^{(\alpha)}(t)$ is an additive uniform white noise with zero mean.  

With this noise term, the dynamics of the system behave similar to a system under continuous stochastic perturbations. Occasionally, these perturbations may drive the system towards the unstable focus, resulting in a decreased rotation frequency, while at other times, the perturbations may direct it toward the unstable limit cycle, leading to an increased rotation frequency (see Fig. \ref{perturbation}(e)). Consequently, both the amplitude and frequency of the oscillation of $ V^{(e)} $ exhibit persistent fluctuations (Fig. \ref{perturbation}(f)). When the rotation frequency of the stable limit cycle approaches the boundary between the $\delta$ and $\theta$ bands (for instance, 3.71 Hz as illustrated in Fig. \ref{perturbation}), these fluctuating frequencies can result in intermittent transitions between states characterized by $\delta$ and $\theta$ rhythms. A statistical analysis of the durations of these two states reveals distributions that are consistent with those observed in Fig. \ref{timeseries} for the QIF model described by Eq. (\ref{QIFModel}) under the same parameters (see Fig. \ref{perturbation}(g) and (h)).

As depicted in Fig. \ref{bifurcation}, the phase space illustrates that the unstable limit cycle is significantly larger and situated further from the stable limit cycle than the unstable focus is from the stable limit cycle. When the effects of fluctuations accumulate, the deviation toward the unstable limit cycle can exceed the deviation toward the unstable focus, resulting in a considerably longer return time. Consequently, the duration of the faster oscillation frequency associated with the $\theta$ rhythm may also be prolonged, as evidenced by the long tail in the distribution of $\theta$ rhythm durations (see Fig. \ref{perturbation}(g)).

\section{Regulation of the power-law exponent}  
In comparison to empirical observations of real EEG data from rats \cite{Wang:2019}, the examples presented in Figures \ref{timeseries} and \ref{perturbation} demonstrate a larger power-law exponent with  $\gamma \approx 4.1$. This indicates that the probability of occurrence for prolonged $\theta$-bursts diminishes at a faster rate than that observed in empirical data. To clarify the underlying factors contributing to this discrepancy, we further investigate the influence of dynamical parameters on the regulation of this power-law behavior. 

\begin{figure}
	\includegraphics[width=0.9\linewidth]{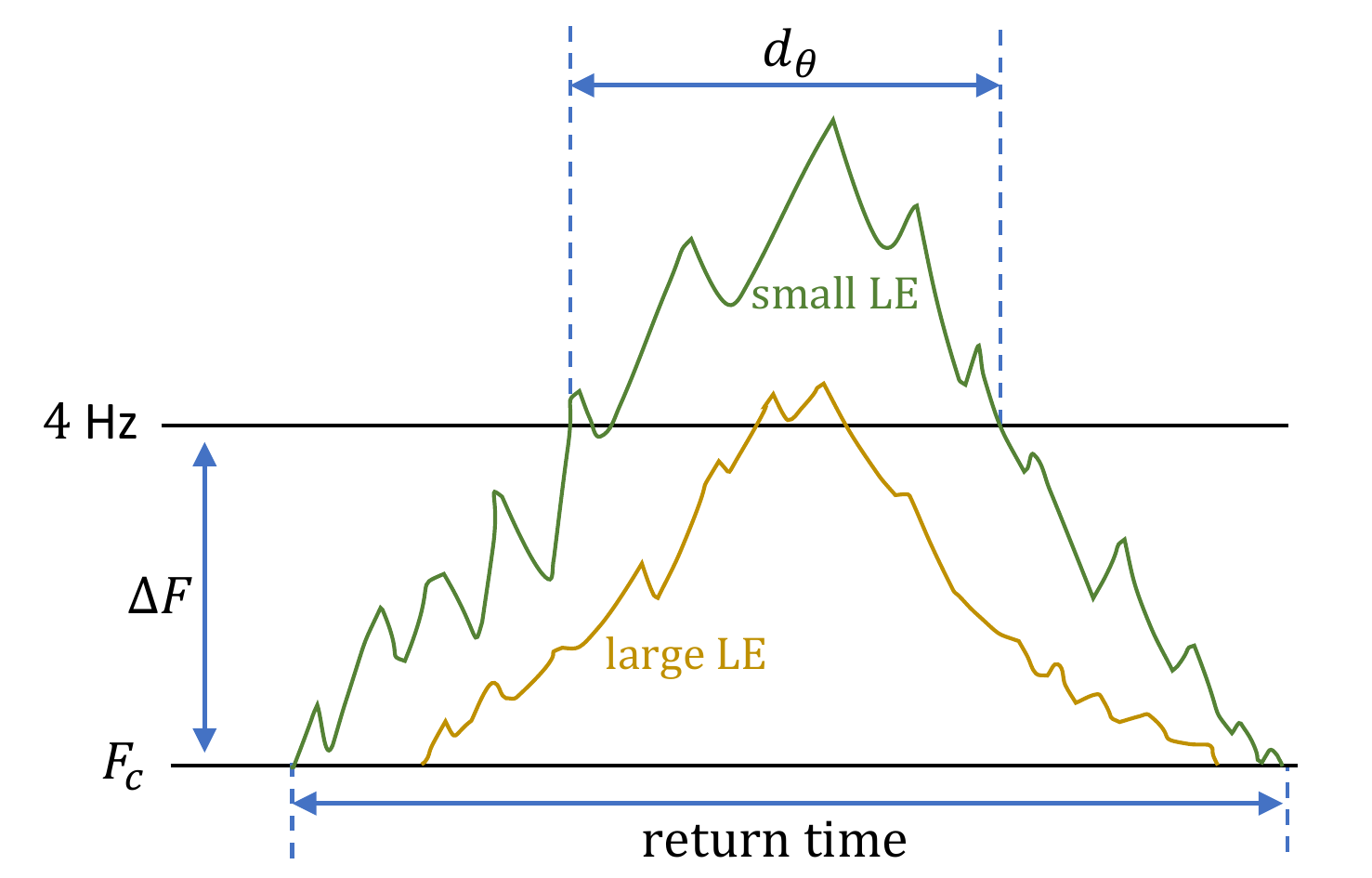} \caption{(color online). A schematic illustration to explain the regulation of $d_{\theta}$. Without fluctuations the system is restricted to the stable limit cycle with an oscillation frequency $F_{c}$. If $F_{c}$ is small, e.g. $F_{c}=3.71$ Hz in Figure \ref{perturbation}, the oscillation of the system is in $\delta$ band. When the fluctuation drives the system towards higher frequency, the effect oscillation frequency starts to increase with fluctuation. If it becomes larger than 4 Hz, the boundary between $\delta$ and $\theta$ bands, the we observe the appearance of $\theta$ rhythm. The time between when the system starts to deviate from $F_{c}$ and when it comes back to $F_{c}$ is defined as the return time. The duration of the system stays above 4 Hz is the duration of observed $\theta$ rhythm. Therefore, there are two factors that may impact $d_{\theta}$. One is $\Delta F$, the difference between $F_{c}$ and 4 Hz, a smaller $\Delta F$ may lead to longer $d_{\theta}$. The second factor is the speed that the system returns to the stable limit cycle, which is governed by the Lyapunov Exponent (LE). A more negative value of LS means a faster return, resulting in a shorter return time and $d_{\theta}$.
		   }
	\label{regulation}
\end{figure}

Figure \ref{regulation} presents a schematic representation illustrating how various factors can influence the observed $d_{\theta}$. In the absence of fluctuations, the system stays at the stable limit cycle with an oscillation frequency $F_{c}$, which is below 4 Hz (the boundary between $\delta$ and $\theta$ bands). When the fluctuation drives the system towards the unstable limit cycle, the oscillation frequency begins to rise. When it is larger enough exceeding the threshold of 4 Hz, the emergence of $\theta$ rhythm is observed. However, due to the stability of the system, it will revert to the stable limit cycle. Therefore, we can define a return time which quantifies the interval between the initial deviation of the oscillation frequency from $F_{c}$ and its subsequent return to $F_{c}$. Furthermore, $d_{\theta}$ represents the duration during which the oscillation frequency remains above 4 Hz. Thus, $d_{\theta}$ is influenced by two primary factors. The first is the Lyapunov Exponents (LEs), which dictate the rate at which the system returns to the stable limit cycle. More negative values of LEs, indicative of a stronger attraction to the limit cycle, correspond to a more rapid return, resulting in a reduced return time and a shorter $d_{\theta}$. The second factor is the deviation of $F_{c}$ from 4 Hz, referred to as $\Delta F$. A smaller $\Delta F$ (indicating a higher $F_{c}$) suggests an increased likelihood of the system remaining within the $\theta$ band (above 4 Hz). Therefore, to achieve an extended duration of $d_{\theta}$ (and a smaller power-law exponent $\gamma$), it is necessary to attain larger LEs that are closer to zero, as well as a higher value of $F_{c}$.

\begin{figure}
	\includegraphics[width=0.9\linewidth]{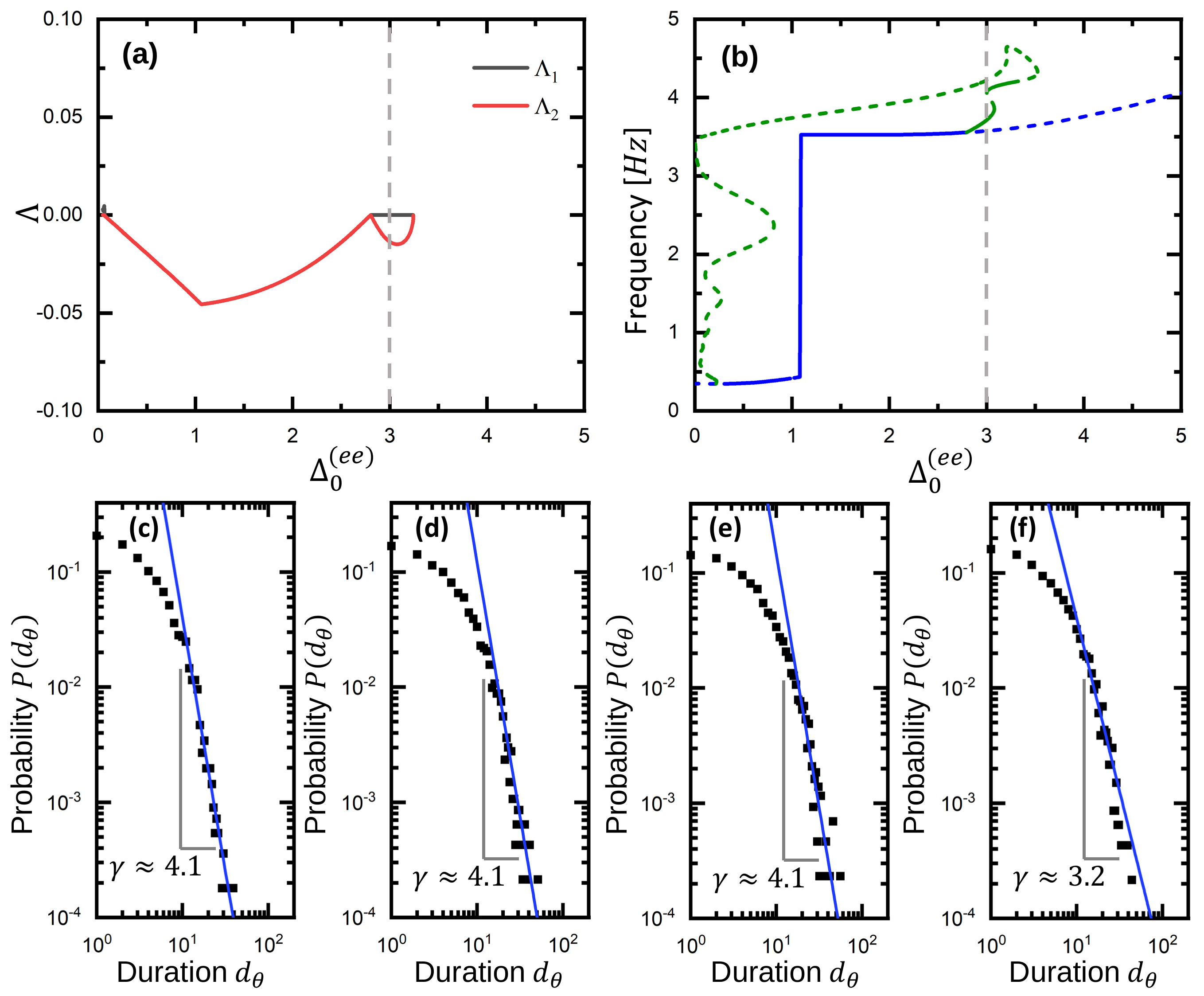} \caption{(color online). (a) The dependence of largest ($\Lambda_1$, black lines) and second largest ($\Lambda_2$, red lines) Lyapunov exponents on $\Delta_{0}^{(ee)}$.  The dependence of the oscillation frequency of $V^{(e)}$ on $\Delta_0^{(ee)}$ obtained from Eq. (\ref{Masstime}). Blue solid (dashed) corresponds to the frequency near the stable (unstable) focus. Green solid (dashed) represent the frequency for the stable (unstable) limit cycle. Bottom panels show the probability density distribution for the duration of $\theta$ rhythm for $K=500$, and (a): $\Delta^{(ee)}_0=2.85$; (b): $\Delta^{(ee)}_0=2.9$; (c): $\Delta^{(ee)}_0=3$; (d): $\Delta^{(ee)}_0=3.2$.	 }
	\label{regulation2}
\end{figure}

Therefore, we calculate the Lyapunov exponents from Eq. (\ref{Massdelta}) (see Appendix for details). For the sake of clarity, we show only the largest Lyapunov exponent ($\Lambda_1$) and the second largest Lyapunov exponent ($\Lambda_2$) within parameter regions that exhibit a stable stationary state. Figure \ref{regulation2}(a) illustrates the variation of $\Lambda_1$ and $\Lambda_2$ as a function of $\Delta_{0}^{(ee)}$. For the case of stable limit cycle, $\Lambda_1$ remains at zero, indicating the neutral stability for the phase of limit cycle. Notably, $\Lambda_2$ approaches its minimum value at $\Delta_0^{(ee)}=3$, indicating that both increases and decreases in $\Delta_0^{(ee)}$ can bring $\Lambda_2$ closer to zero. This suggests a slower return to the stable limit cycle following perturbations, which is advantageous for the sustained presence of long-lasting $\theta$ rhythms. Furthermore, Figure \ref{regulation2}(b) depicts the relationship between rotation frequency of the limit cycle and $\Delta_{0}^{(ee)}$ (represented by green lines). Close to the region where $\Delta_{0}^{(ee)}=3$, the rotation frequency of the stable limit cycle demonstrates a consistent increase with higher values of $\Delta^{(ee)}_0$. Consequently, an increase in $\Delta_{0}^{(ee)}$ is favorable for the emergence of a long-lasting $\theta$ rhythm, as illustrated schematically in Figure \ref{regulation}. 

When considering the combined effects of variations in Lyapunov exponents and limit cycle frequency, a decrease in $\Delta_{0}^{(ee)}$ from 3 appears to have a negligible impact on the power-law exponent, as the opposing effects tend to offset one another. Conversely, an increase in $\Delta_{0}^{(ee)}$ yields beneficial changes in both Lyapunov exponents and limit cycle frequency, which collectively promote the emergence of a long-lasting $\theta$ band, potentially resulting in a reduction of the observed power-law exponent $\gamma$. This assertion is corroborated by the simulation results presented in the lower panels of Figure \ref{regulation2}. Specifically, the values of $\gamma$ for $\Delta^{(ee)}_0=2.85$, $\Delta^{(ee)}_0=2.9$, and $\Delta^{(ee)}_0=3$ remain consistent (as shown in panels (c) to (e)), while a decrease to $\gamma \approx3.2$ is observed for $\Delta^{(ee)}_0=3.2$ (as depicted in panel (f)). 

\begin{figure}
	\includegraphics[width=0.9\linewidth]{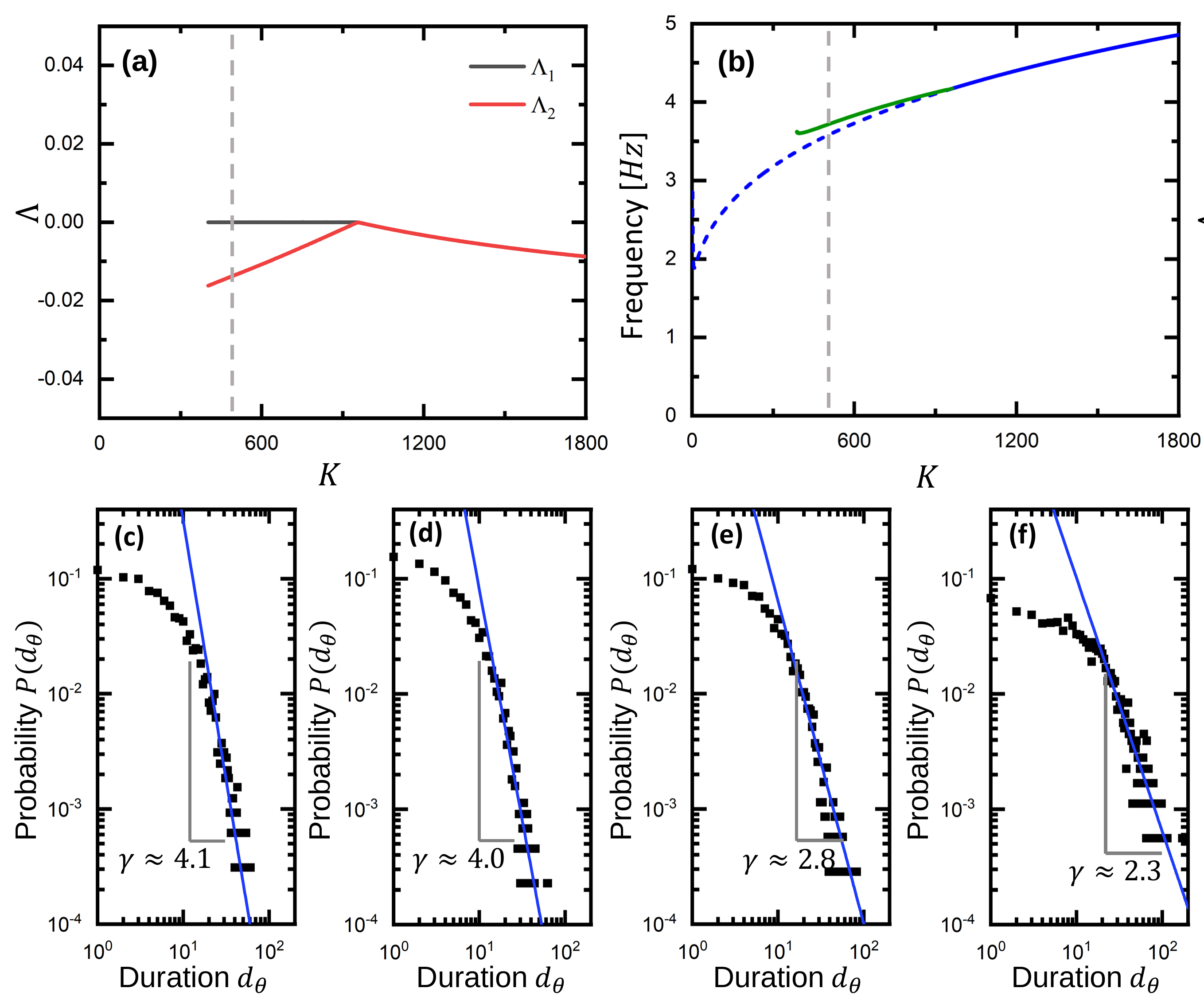} \caption{(color online). Upper panels represent the probability density distribution for the duration of $\theta$ rhythm for $K=500$, and (a): $\Delta^{(ee)}_0=2.85$; (b): $\Delta^{(ee)}_0=2.9$; (c): $\Delta^{(ee)}_0=2.95$; (d): $\Delta^{(ee)}_0=3$. The power-law exponent does not change. Bottom panels represent the probability density distribution for the duration of $\theta$ rhythm for $\Delta^{(ee)}_0=3$, and (e): $K=500$; (f): $K=600$; (g): $K=700$; (h): $K=800$. The power-law exponent decreases from 4.1 to 2.3. }
	\label{regulation3}
\end{figure}

In Figure \ref{regulation3}, we illustrate the impact of varying the parameter $K$. Both $\Lambda_2$ and the rotation frequency of the stable limit cycle exhibit a consistent increase as $K$ rises (see Figures \ref{regulation3}(a) and (b)). These effects are advantageous for the sustained presence of the long-lasting $\theta$ rhythm, leading to a reduction in the value of $\gamma$ from 4.1 to 2.3 as network connectivity increases from $K=500$ to $K=800$ ( Figures \ref{regulation3}(c) to (f)). Notably, for the parameters $K=800$ and $\Delta^{(ee)}_0=3$, a power-law exponent of $\gamma \approx 2.3$ is achieved, which is in close agreement with empirical observations of healthy sleep patterns in rats \cite{Wang:2019,Huo:2024,Lombardi:2020}.

\section{Conclusion}

Spontaneous transitions among various states of cortical dynamics along with power-law behaviors have been empirically examined across multiple levels, ranging from in vitro neuronal studies and cortical slice cultures \cite{Pasquale:2008,Beggs:2003} to human electroencephalography (EEG) and functional magnetic resonance imaging (fMRI) \cite{Hansen:2001,Palva:2013,Marinazzo:2014,Tagliazucchi:2014}, as well as the dynamics associated with sleep-stage and arousal transitions \cite{Lo:2002,Lo:2013,Sorribes:2013}. Models grounded in statistical physics indicate that such intermittent behaviors are typically associated with the critical dynamics of nonequilibrium systems \cite{Lombardi:2023,Dvir:2018,Fraiman:2009,Stramaglia:2017,Miranda:1991}.

In this paper, we focus on the intermittent transitions between $\theta$ and $\delta$ bursts observed on EEG data recorded from rats.  Utilizing a model of sparsely coupled excitatory and inhibitory populations of quadratic integrate-and-fire (QIF) neurons, we elucidate how these intermittent transitions and the associated power-law distribution arise from intrinsic fluctuations within a finite-sized system situated near a critical point, specifically the Hopf bifurcation point. Furthermore, we analyze the influence of network connectivity parameters on the stability (measured by the Lyapunov Exponent) and the oscillation frequency of the limit cycle, which leads to power-law distributions with varying exponents $\gamma$. Our findings, from the viewpoint of the basic dynamics of neuronal networks, offer a potential dynamical mechanism to account for the observed intermittent transitions between cortical rhythms with power-law distributions during sleep.

Evidence suggests that variations in critical behavior and the associated power-law exponents may indicate pathological conditions \cite{Zimmern:2020}. For instance, rat models exhibiting sleep disorders demonstrates differing power-law exponents in the duration distribution of $\theta$ bursts. Lesions in the wake-promoting locus coeruleus (LC) result in prolonged sleep (hypersomnia) and a corresponding decrease in the power-law exponent for $\theta$ burst duration \cite{Huo:2024}. Conversely, lesions in the sleep-promoting ventrolateral preoptic nucleus (VLPO) lead to reduced sleep (insomnia) and an increase in the power-law exponent for $\theta$ burst duration \cite{Lombardi:2020}. At a macroscopic level, alterations in the power-law exponent are also observed in the wake/arousal duration distribution under various conditions. For example, sleep apnea is associated with an increase in the power-law exponent \cite{Lo:2013}. Additionally, higher body temperatures have been shown to elevate the power-law exponent for wake/arousal duration distribution, which may explain the heightened risk of sudden infant death syndrome associated with increased ambient temperatures \cite{Dvir:2018}. Given the relationship between sleep stages and cortical rhythms \cite{Rechtschaffen:1968}, these changes in the power-law exponent for wake duration also reflect alterations in the duration of corresponding cortical rhythms at the neuronal network level. Although our model comprises only one excitatory and one inhibitory neuronal population, our findings still suggest a potential mechanism by which changes in neuronal network connectivity can lead to the observed variations in power-law exponents associated with sleep disorders. In a broader context, pathological conditions may contribute to sleep disorders while simultaneously affecting cortical connectivity. For instance, Parkinson's disease frequently results in sleep disturbances \cite{Schrempf:2014,Comella:2007}, including both insomnia and hypersomnia. Furthermore, Parkinson's disease is known to induce changes in cortical connectivity \cite{Cerasa:2016}. Consequently, our results provide a mechanism that links the altered connectivity associated with Parkinson's disease to changes in neuronal network activity and sleep disorders at the macroscopic level.

\section{ACKNOWLEDGMENTS}

This work was supported by the NNSF of China (grant No. 12105117 and 12165016), Guangdong Basic and Applied Basic Research Foundation (grant No. 2024B1515020079 and 2022A1515010523), Shenzhen Science and Technology Program (JCYJ20230807120805010).

\section{Appendix} 
\subsection{Lyapunov Analysis}  

To analyze the linear stability of generic solutions of Equation (\ref{Masstime}), we estimate the corresponding Lyapunov spectrum (LS) $\{\Lambda_{k}\}$. This can be done by considering the time evolution of the tangent vector $\delta = \{\delta R^{(\alpha)}, \delta V^{(\alpha)}, \delta Q_{2}^{(\alpha)}, \delta P_{2}^{(\alpha)}\}$, , that is ruled by the linearization of the Equation (\ref{Masstime}), namely
	\begin{eqnarray}
		\tau_m \delta\dot{R}^{(\alpha)}&=&2R^{(\alpha)}\delta V^{(\alpha)}+(2V^{(\alpha)}+\Delta^{(\alpha\alpha)}_0|g^{(\alpha\alpha)}_{0}|\pi^{-1})\delta R^{(\alpha)} \nonumber\\
		&+&\delta P_{2}^{(\alpha)}(\tau_m\pi)^{-1}, \nonumber\\
		\tau_m \delta \dot{V}^{(\alpha)}&=&2V^{(\alpha)}\delta V^{(\alpha)}-2(\pi\tau_m)^{2} R^{(\alpha)}\delta R^{(\alpha)} \nonumber\\
		&+&\sqrt{K}\tau_m(g^{(\alpha\alpha)}_{0}\delta R^{(\alpha)} +g^{(\alpha\beta)}_{0} \delta R^{(\beta)})+\delta Q_{2}^{(\alpha)},\nonumber\\
		\tau_m\delta \dot{Q}_{2}^{(\alpha)}&=&2\mathcal{N_{R}}^{'}+4(Q_2^{(\alpha)}\delta V^{(\alpha)}+V^{(\alpha)}\delta Q_2^{(\alpha)}  \nonumber\\
		&-&\pi \tau_m P_2^{(\alpha)}  \delta R^{(\alpha)}-\pi\tau_m  R^{(\alpha)}\delta P_2^{(\alpha)}),\nonumber\\
		\tau_m\delta \dot{P}_{2}^{(\alpha)}&=&2\mathcal{N_{I}}^{'}+4(P_2^{(\alpha)}\delta V^{(\alpha)}+V^{(\alpha)}\delta P_2^{(\alpha)} \nonumber\\
		&+&\pi \tau_m   Q_2^{(\alpha)} \delta  R^{(\alpha)}+\pi \tau_m  R^{(\alpha)}\delta Q_2^{(\alpha)}),
		\label{Massdelta}
	\end{eqnarray}
	with $\mathcal{N_R}^{'}=\frac{(g^{(\alpha\alpha)}_{0})^2 \tau_m \delta R^{(\alpha)}}{2K}+\frac{(g^{(\alpha\beta)}_{0})^{2}\tau_m \delta R^{(\beta)}}{2K}$,  $\mathcal{N_I}^{'}=-\frac{(g^{(\alpha\alpha)}_{0})^2 \Delta^{(\alpha\alpha)}_0 \tau_m \delta R^{(\alpha)}}{2K}=-\Delta^{(\alpha\alpha)}_0 \mathcal{N_R}^{'}$, and $\Delta_0^{(\alpha\beta)}=\Delta_0^{(\beta\alpha)}=0$. In this case, the LS is composed of eight Lyapunov exponents(LEs) $\{\lambda_{k}\}$ with $k = 1,\dots, 8$, which quantify the average growth rates of infinitesimal perturbations along the orthogonal manifolds. The LEs can be estimated as follows:
	\begin{align}
		\Lambda_{k} = \lim_{t \to \infty}\frac{1}{t}\log \frac{|\delta_{k}(t)|}{|\delta_{k}(0)|},
		\label{FPF9}
	\end{align}
	where the tangent vectors $\delta_{k}$ are maintained ortho-normal during the time evolution by employing a standard technique.  The autonomous system will be chaotic for $\Lambda_1 > 0$, while a periodic (two frequency quasi-periodic)
	dynamics will be characterized by $\Lambda_1 = 0$ ($\Lambda_1 = \Lambda_2 = 0$) and a fixed point by $\Lambda_1 < 0$.
	In order to estimate the LS for the neural mass model, we have integrated the direct and tangent space evolution with a RungeKutta $4th$ order integration scheme with $dt = 0.01$ ms, for a duration of $200$ s, after discarding a transient of $10$ s.

\end{document}